\documentstyle[psfig]{mn}
\newcommand{\Msolar}{\mbox{${\rm M}_{\odot}$}}

\begin{document}
\title[WD\,1953$-$011] {WD\,1953$-$011 -- a magnetic white dwarf with peculiar
field structure.}
\author[P.F.L. Maxted et~al.]
       { P. F. L. Maxted$^1$, L. Ferrario$^2$, T. R. Marsh$^1$,
        and D.~T. Wickramasinghe $^2$  \\
        $^1$University of Southampton, Department of Physics \& Astronomy,
        Highfield, Southampton, S017 1BJ, UK \\
        $^2~$Department of Mathematics and Astrophysical Theory Centre,
Australian National University, Canberra, ACT 0200, Australia }
\date{Accepted 1999 
      Received 1999 }

\date{Accepted. Received ; in original form}
\maketitle
\label{firstpage}

\begin{abstract}
  
  We present H$_{\alpha}$ spectra of the magnetic white dwarf star
WD\,1953$-$011 which confirm the presence of the broad Zeeman components
corresponding to a field strength $\sim 500$\,kG found by Maxted \& Marsh
(1999). We also find that the line profile is variable over a timescale of a
day or less.  The core of the H$_{\alpha}$ line also shows a narrow Zeeman
triplet corresponding to a field strength of $\sim 100$\,kG which appears to
be almost constant in shape. These observations suggest that the magnetic
field on WD\,1953$-$011 has a complex structure and that the star has a
rotational period of hours or days which causes the observed variability of
the spectra. We argue that neither an offset dipole model nor a double-dipole
model are sufficient to explain our observations.  Instead, we propose a two
component model consisting of a high field region of magnetic field strength
$\sim 500$\,kG covering about 10\% of the surface area of the star
superimposed on an underlying dipolar field of mean field strength $\sim
70$\,kG.  Radial velocity measurements of the narrow Zeeman triplet show that
the radial velocity is constant to within a few km/s so this star is unlikely
to be a close binary.

\end{abstract}

\section{Introduction}

There are now about 60 magnetic white dwarfs known with fields in the range
of 0.1\,MG to 1000\,MG (Wickramasinghe \& Ferrario 2000). The distribution of
field strength above 1\,MG is approximately consistent with equal numbers in
each decade of field (Schmidt \& Smith 1995). Significant progress has been
made in recent years in the interpretation of the spectra of magnetic white
dwarfs  and through such studies a detailed picture is emerging on their
surface field structure.  These in turn can be related, through models of
field decay and evolution, to the properties of their progenitors which are
likely to be the magnetic Ap and Bp stars (Angel et al. 1981).

WD\,1953$-$011 is a white dwarf star that has a variety of aliases (G\,92$-$40,
GJ\,772, EG\,135, LHS\,3501, L\,997$-$21) which reflects a history of
observations spanning several decades. It is one of the closest white dwarfs
to the Earth ($d=12$pc;  Harrington \& Dahn 1980) but is quite faint because
it is rather cool (V=13.7, T$_{\rm eff}$=7932\,K; Bragaglia et~al. 1995).
Measurements of the magnetic field strength of WD\,1953$-$011 give a confusing
picture. Schmidt \& Smith (1995) used circular spectropolarimetry to measure a
mean longitudinal field strength of $-15.1\pm6.6$\,kG. Koester et~al. (1998)
observed the core of the H$_{\alpha}$ line at high resolution and found that 
the spectrum shows a narrow Zeeman triplet corresponding to a mean surface
field strength of $93\pm5$\,kG. Maxted \& Marsh (1999) obtained a single
spectrum of WD\,1953$-$011 with lower resolution but with better
wavelength coverage which showed broad depressions in the wings of the
H$_{\alpha}$ line which, if they are interpreted as being Zeeman components,
correspond to a mean surface field strength of $\sim 0.5$\,MG. It is clear
that WD\,1953$-$011 is a magnetic white dwarf star but these observations
suggest that its magnetic field is unlikely to have a simple structure.

In this letter, we present a series of spectra of WD\,1953$-$011 which confirm
the presence of the broad Zeeman components of H$_{\alpha}$ observed by Maxted \&
Marsh (1999) and which show that the strength of these broad features is
variable. We argue that simple dipole models of the magnetic field are not able
to explain these features but present a preliminary model of the field
structure in which the broad Zeeman features are due to a ``magnetic spot'' on
the surface of the star.

\section{Observations}

We observed  WD\,1953$-$011 on the nights 1999 August 1\,--\,3 with the
Anglo-Australian Telescope at Siding Spring Observatory, Australia. The
instrument and data reduction process are identical to those described
in Maxted \& Marsh (1999) with the exception of the detector, which gave
slightly better sampling of the spectra in the 1999 spectra. We  obtained 12
spectra of WD\,1953$-$011 around the H$_{\alpha}$ line with  a resolution of
0.74\AA.  Details of these spectra and the original spectrum of Maxted \&
Marsh are given in Table~\ref{LogTable}. The numbers assigned to the spectra
therein will be used throughout this paper when we need to refer to individual
spectra.

\section{Analysis}

\begin{table}
\caption{\label{LogTable} A log of our observation of WD\,1953$-$011. T$_{\rm
exp}$ is the exposure time in seconds, UT is the time of mid-exposure , RV
is the radial velocity of the narrow Zeeman triplet in km/s and
$\delta\lambda$ in the splitting between the $\sigma_+$  and
$\sigma_-$ components of the narrow Zeeman  triplet in km/s.}
\begin{tabular}{rrrrrr}
\multicolumn{1}{l}{N} &
\multicolumn{1}{l}{T$_{\rm exp}$} &
\multicolumn{1}{l}{Date} &
\multicolumn{1}{l}{UT} &
\multicolumn{1}{l}{RV} &
\multicolumn{1}{l}{$\delta\lambda$}\\
 1 &  600 & 16 Aug 1997 & 10:55:40 & 59.4 $\pm$ 3.4 &  91 $\pm$ 5 \\
 2 &  600 & 01 Aug 1999 & 10:45:10 & 59.9 $\pm$ 3.5 &  93 $\pm$ 4 \\
 3 &  600 & 01 Aug 1999 & 10:55:30 & 61.5 $\pm$ 2.8 & 108 $\pm$ 4 \\
 4 &  600 & 01 Aug 1999 & 11:05:52 & 58.5 $\pm$ 2.8 &  93 $\pm$ 4 \\
 5 & 1800 & 01 Aug 1999 & 13:25:27 & 59.6 $\pm$ 1.5 &  93 $\pm$ 2 \\
 6 & 1800 & 02 Aug 1999 & 10:58:18 & 60.1 $\pm$ 1.0 &  83 $\pm$ 1 \\
 7 & 1800 & 02 Aug 1999 & 13:35:14 & 60.7 $\pm$ 1.2 &  80 $\pm$ 2 \\
 8 & 1800 & 02 Aug 1999 & 14:32:41 & 58.5 $\pm$ 1.1 &  83 $\pm$ 1 \\
 9 & 1200 & 03 Aug 1999 & 10:39:03 & 59.5 $\pm$ 2.2 &  92 $\pm$ 3 \\
10 & 1200 & 03 Aug 1999 & 11:00:02 & 56.6 $\pm$ 1.8 &  87 $\pm$ 2 \\
11 & 1200 & 03 Aug 1999 & 11:21:15 & 58.4 $\pm$ 1.7 &  84 $\pm$ 2 \\
12 & 1200 & 03 Aug 1999 & 11:42:21 & 58.5 $\pm$ 1.7 &  84 $\pm$ 2 \\
13 & 1200 & 03 Aug 1999 & 12:03:39 & 60.8 $\pm$ 1.8 &  83 $\pm$ 2 \\
\end{tabular}
\end{table}

\subsection{General appearance.}
\begin{figure} 
\caption{\label{SpecFig} Spectra of WD\,1953$-$011. The spectra are offset
vertically for clarity and are plotted in a temporal sequence from bottom
(earliest) to top (latest).}
\psfig{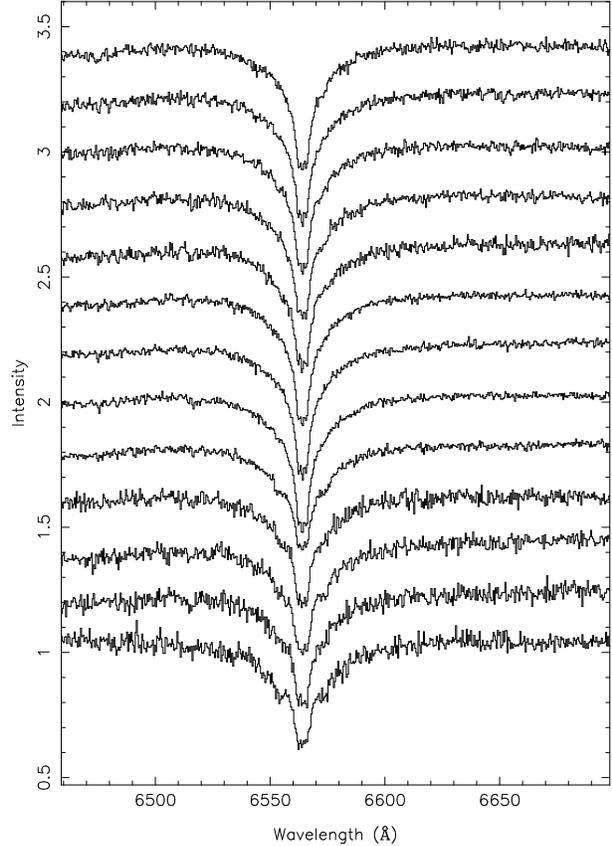} 
\end{figure}

All the spectra are shown in Fig.~\ref{SpecFig} following normalization using
a linear fit to the continuum either side of the H$_{\alpha}$ line. The narrow
Zeeman triplet in the core of the H$_{\alpha}$ line observed by Koester et~al.
(1998) is seen in most spectra.  It is clear that the spectrum is variable
particularly with regard to the overall depth of the H$_{\alpha}$ line and the
strength of the broad features at 6554\AA\ and 6576\AA. There is also some
indication that the splitting of the narrow Zeeman components varies by about
10\% (see section \ref{RVSect}). No period can be assigned to the variability
from our data, but the timescale of the variations is hours or days.

\subsection{\label{RVSect} Radial velocity.}
We first suspected that the presence of two distinct magnetic fields might be
explained by WD\,1953$-$011 being a binary star so we measured the radial
velocities given in Table~\ref{LogTable} from the narrow Zeeman triplet in the
core of H$_{\alpha}$ using a least-squares fit of model profile formed from
several Gaussian profiles. The radial velocity is constant to within a few
km/s. A more careful analysis shows that we would have detected over 90\% of
binaries with an orbital period shorter than 20d. The mean radial velocity
is 59.7$\pm$0.2\,km/s, which is typical for white dwarfs. The splitting
between the  $\sigma_+$ and $\sigma_-$ components in the narrow Zeeman
triplet was included as a variable in the fit and is also given in
Table~\ref{LogTable}. It is clear that the splitting is variable by about
10\%, though no period can be identified from these data.

\section{The structure of the magnetic field.}

While the radial velocity measurements are not conclusive concerning
binarity, the observational constraint, namely that the satellite features
corresponding to the high field component are only present at some phases,
strongly supports a single star interpretation. 

The Zeeman spectra of most white dwarfs can be modelled by  a centered or
off-centered dipole field distribution so we begin by investigating if such a
model can be constructed to explain the spectra of WD\,1953$-$011.

The splitting between the mean $\sigma_+$ and $\sigma_-$ components in the
narrow Zeeman triplet corresponds to $\delta \lambda= 1.9$\AA.
%
%
The spectra can be modelled in the standard way (see Martin \& Wickramasinghe
1979), but since Stark broadening dominates over Zeeman splitting at low
fields, the model fits are not very sensitive to the field geometry. If we
assume a centered dipole field distribution, those spectra where the triplet
is clearly seen (6\,--\,13) can be modelled with a dipolar field strength
$B_d= 100$\,kG (corresponding to an average surface field strength of 70~kG)
and a viewing angle $i=50^{\circ} \pm 20^{\circ}$ measured from the dipole
axis.

\begin{figure} 
\caption{\label{FitFig} A possible two-component model for WD\,1953$-$011
covering a whole spin cycle. The model corresponds to a centred dipolar field
with $B_d= 100$\,kG and a spot with a field of $490\pm 60$\,kG covering 10\%
of the stellar surface. The two sets of data overlapped to the models
correspond to spectrum 1 (bottom) and spectrum 6 (top) of the time series
shown in Figure 1.}  
\psfig{file=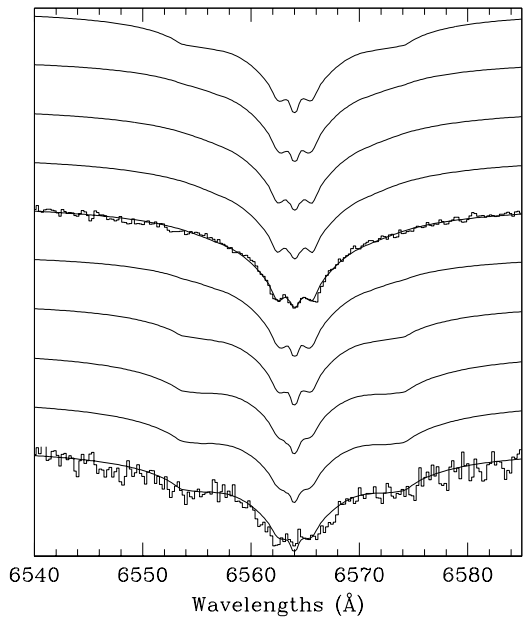,width=0.6\textwidth}
\end{figure} 

If we identify the weak satellite features seen in spectra 1\,--\,5 at
6554\AA\ and 6576\AA\ with the $\sigma$ components of $H_{\alpha}$ the
measured separation between these components $\delta \lambda= 20$\AA\ implies
a mean field strength of $B_{high}\approx 500$\,kG, about 8 times higher
than the field strength deduced from the narrow Zeeman triplet. The centered
dipole model that was used to model the low field Zeeman triplet cannot be
used to model the higher field $\sigma$ components simply by changing the
viewing aspect because a centered dipole model only allows a field spread of a
factor 2, which is much lower than that observed.

A larger field spread can be obtained quite simply by offsetting the dipole
from the centre of the star. In this case the field strengths
$B_{p1}$ and $B_{p2}$ at the opposite poles are given by
\[ \frac{B_{p1}}{B_{p2}} = \left(\frac{1-d}{1+d}\right)^3\]
where the dipole is displaced by a fraction $d$ of
the stellar radius in the direction of the dipole axis.

In the extreme case where the low field and the high field phases correspond
to direct viewing of the opposite poles and the maximum field contrast is
provided, the observations indicate $d \approx 0.35$.  If we relax this
assumption, even higher values of $d$ would be required to achieve the
observed field contrast. However, attempts at such a model fit show that
although the weaker field Zeeman triplet is well fitted at the phase when the
weaker pole is viewed most directly, at the opposite phase, when the stronger
pole is viewed most directly, the broadening due to field spread is far too
severe to match the widths of the high field $\sigma$ components.  This is a
property of all offset models - the hemisphere with the weaker pole has a more
uniform field distribution than the other hemisphere, and the contrast becomes
extreme for highly offset dipoles (Martin \& Wickramasinghe 1984). We must
therefore conclude that global field structures of the offset dipole type
cannot model the observations of WD\,1953$-$011. Nor it is possible to
construct a double-dipole model (a strongly offset dipole superimposed on a
centred dipole) similar to that proposed for PG~1031+254 (Schmidt et~al.
1986; Latter et~al. 1987).

It is a remarkable observational fact that even at the phases where the high
field components are seen, the low field Zeeman triplet appears to be present
determining the width of the line core. Furthermore, whenever the high field
component is seen, it appears to occur at almost the same wavelength
corresponding essentially to one field strength.  The most likely explanation
is that there is a nearly uniform region of enhanced field strength which
comes into view only at certain rotational phases, but that we are seeing
contributions from the low field region at all phases.
 
We have constructed such a model, and confirmed that it is indeed possible to
fit the data at both the high and low field phases rather well. The model
consists of two components: a high field region with a nearly uniform field of
strength $490\pm 60$\,kG covering 10\% of the surface area of the star, and a
second lower field component which is modelled by a centered dipolar field
distribution with $B_d=100$\,kG. The viewing geometry is such that the spot is
visible for approximately half of the spin cycle. Since we do not have a
complete spin cycle, the phasing and the relative importance of the two
components are not strongly constrained and the model shown in Figure 2 cannot
be claimed to be unique.  However, it encompasses the basic characteristics
that are required to fit the observations.

Any plausible model for the field structure must satisfy the important
constraint that the total magnetic flux over the surface of the star must be
zero.  The double-dipole models clearly satisfy this constrain, but they do
not fit the data.  For the high field region with a nearly uniform field, the
zero flux condition could imply a magnetic spot type field enhancement (that
is, a closed magnetic loop), but this appears unlikely, since there is no
evidence for late-type star magnetic activity in white dwarfs.  A more likely
explanation is that the nearly uniform field region that we appear to require
is in fact a part of a global field structure which diverges from that of
either a centred or an off-centred dipole or a combination of both and that
the zero magnetic flux condition is satisfied in a global sense over the
entire star.  We expect that when phase-resolved spectropolarimetric data
become available over a full rotational cycle, a fit to the data using a
multi-polar expansion would yield a field structure of the type mentioned
above. 

\section{Discussion and conclusions}

Studies of the variable spectra in rotating magnetic white dwarfs have posed
two questions, namely (i) the origin of the rapid rotation, and (ii) the
apparently peculiar magnetic field structures. Rapid rotation in a magnetic
white dwarf may be the result either of spin-up in a binary or of a binary
merger. Thus, systems like Feige~7 (Achilleos et~al. 1992) or PG~1031+234
(Schmidt et al. 1986; Latter et~al. 1987) may be extinct AM Herculis
variables with very low-mass and hitherto undetected companions. On the other
hand, the high mass and rapid rotation deduced for RE~J0317$-$853
(=EUVE~J0317$-$855;  Barstow et al. 1995) is strongly suggestive of spin up
during a merger (Ferrario et~al. 1997). The situation may indeed be similar to
what is found for neutron stars where rapid rotation in an old system is taken
as evidence of spin-up in a binary.

However, the origin and the nature of the complex field structures seen in
magnetic white dwarfs remains a mystery (Wickramasinghe \& Ferrario 2000).
This question should be viewed in the context of the growing evidence that the
isolated magnetic white dwarfs tend as a class also to have a higher mass than
their non-magnetic magnetic or weakly magnetic counterparts implying a
different post main sequence evolutionary history. Is the complexity
correlated with field strength, and how is it related to the mass of the white
dwarf and the nature of the progenitor star and its evolution? So far, stars
with non-standard field structures have also exhibited high field strengths
and almost all show rapid rotation. However, our observations of
WD\,1953$-$011 have provided evidence of what appears to be one of the most
peculiar field structures in any isolated magnetic white dwarf, but this time
in a very low field object. Future observations of WD\,1953$-$011 which would
reveal the structure of the magnetic field in more detail include further
spectroscopy, perhaps at other wavelengths, and circular spectropolarimetry.
The value of observations at far-UV wavelengths is demonstrated  by the study
of RE~J0317$-$853 (Burleigh, Jordan \& Schweizer 1999), an extreme magnetic
white dwarf star with a rotation period of only 725s and an effective
temperature of about 40,000\,K. They were able to model the surface
distribution of the magnetic field strength using an expansion into spherical
harmonics and fit their phase-resolved spectra of the Ly-$\alpha$ line and the
``forbidden'' $1s_0\rightarrow 2s_0$ line. The range of field strengths they
measured is 180--800\,MG. Circular spectropolarimetry of RE~J0317$-$853
(Ferrario et~al. 1997; Jordan \& Burleigh 1999) shows the highest level of
polarization (22\,percent) measured to-date on an isolated white dwarf star.
RE~J0317$-$853 is much hotter than WD\,1953-011 though they  are similar in
their optical brightness.  However, the rotation period of WD\,1953$-$011 is
much longer than the 725s rotation period of RE~J0317$-$853, so it should be
possible to study the magnetic field in greater detail.

The progenitors of the vast majority of the magnetic white dwarfs are likely
to be the magnetic Ap and Bp stars. The mass of WD\,1953$-$011
(0.844$\pm$0.035\Msolar ; Bragaglia et~al. 1995) is typical for magnetic white
dwarfs (Fabrika \& Valyavin 1999) and suggests that it is the remnant of just
such a star. A fossil origin for the fields is therefore likely. Some Ap stars
show complicated field structures with evidence of strong quadrupolar
components and spot type field structures (Landstreet 1980; Mestel 1999). The
simplest theories of the Ohmic decay of fossil fields predict that the higher
order modes decay fastest (Wendell et al. 1987). Strong quadrupolar or higher
order components are therefore not expected in the cooler stars. Recently,
Muslimov et al. (1995) have re-investigated this problem allowing for an
additional effect: the Hall drift. They have shown that through the non-linear
coupling of toroidal and poloidal modes, it is possible to have, under certain
circumstances, an enhancement of the quadrupolar component relative to the
dipolar component as the white dwarf cools. The field structure in
WD\,1953$-$011 would appear to suggest that additional effects, such as Hall
currents, may indeed play a r\^{o}le in the evolution of magnetic fields in
white dwarfs.

\section*{Acknowledgements}
 PFLM was supported by a PPARC post-doctoral grant.  We would like to thank
the anonymous referee for their constructive comments.

\label{lastpage}


\begin{thebibliography}{}
\bibitem{2} Achilleos N., Wickramasinghe D.T.,
  Liebert J., Saffer R., Grauer A.D., 1992, ApJ, 396, 273
\bibitem{3} Angel J.R.P., Borra E.F., Landstreet J.D., 1981, ApJS, 45, 457
\bibitem{4}Barstow M.A., Jordan S., O'Donoghue D., Burleigh M.R., Napiwotzki R.,
 Harrop-Allin M.K., 1995, MNRAS, 277, 971
\bibitem{z} Bragaglia A., Renzini A., Bergeron P., 1995, ApJ, 443, 735
\bibitem{x} Burleigh M.R., Jordan S., Schweizer W., 1999, ApJ, 510, L37 
\bibitem{q} Fabrika S., Valyavin G., 1999, Proceedings of the 11th European
Workshop on White Dwarfs, ASP conference series 169, Eds. J.-E. Solheim and E.
G. Meistas, p. 214
\bibitem{5}Ferrario L., Vennes S., Wickramasinghe D.T., Bailey J.A., 
Christian D., 1997, MNRAS, 292, 205
\bibitem{6}Harrington R.S., Dahn C., 1980, AJ, 85, 454
\bibitem{y}Jordan S., Burleigh M.R., 1999, Proceedings of the 11th European
Workshop
on White Dwarfs, ASP conference series 169, Eds. J.-E. Solheim and E. G.
Meistas, p. 235.
\bibitem{9}Koester D., Dreizler S., Weidemann V., Allard N.F., 1998, A\&A, 338, 612
\bibitem{10}Landstreet J.D., 1980, AJ, 85, 611
\bibitem{11}Latter W.B., Schmidt G. D., Green R.F., 1987, ApJ, 320, 308
\bibitem{12}Martin B., Wickramasinghe D.T., 1979, MNRAS, 189, 883
\bibitem{13}Martin B., Wickramasinghe D.T., 1984, MNRAS, 206, 407
\bibitem{a}Maxted P.F.L., Marsh T.R., 1999, MNRAS, 307, 122.
\bibitem{b}Mestel L., 1999, in ``Stellar Magnetism'', International Series of 
Monographs on Physics, Oxford University Press
\bibitem{c}Muslimov A.G., Van Horn H.M., Wood M.A., 1995, ApJ, 442, 758
\bibitem{e}Schmidt G.D., West S.C., Liebert J., Green R.F., Stockman H.S.,
1986, ApJ, 309, 218
\bibitem{f}Schmidt G.D., Smith P.S., 1995, ApJ, 448, 305  
\bibitem{g}Wendell C.E., Van Horn H.M., Sargent D., 1987, ApJ, 313, 284
\bibitem{h}Wickramasinghe D.T., Ferrario L., 2000, PASP, In press.
\end{thebibliography}
\end{document}